\documentclass[conference, a4paper]{IEEEtran}
\IEEEoverridecommandlockouts
\usepackage{cite}
\usepackage{amsmath,amssymb,amsfonts}

\usepackage{algorithm}
\usepackage{graphicx}
\usepackage{textcomp}
\usepackage{xcolor}
\usepackage{geometry}
\usepackage{float}
\usepackage{algorithmic}

\usepackage[hidelinks]{hyperref}

\def\BibTeX{{\rm B\kern-.05em{\sc i\kern-.025em b}\kern-.08em
    T\kern-.1667em\lower.7ex\hbox{E}\kern-.125emX}}
\begin{document}

\title{Stable-by-Design Neural Network-Based LPV State-Space Models for System Identification}

\author{\IEEEauthorblockN{Ahmet Eren Sertbaş, Tufan Kumbasar}
\IEEEauthorblockA{\textit{Artificial Intelligence and Intelligent Systems Laboratory} \\
\textit{Istanbul Technical University}\\
Istanbul, Türkiye \\
sertbas20@itu.edu.tr, kumbasart@itu.edu.tr}

}

\maketitle

\begin{abstract}
Accurate modeling of nonlinear systems is essential for reliable control, yet conventional identification methods often struggle to capture latent dynamics while maintaining stability. We propose a \textit{stable-by-design LPV neural network-based state-space} (NN-SS) model that simultaneously learns latent states and internal scheduling variables directly from data. The state-transition matrix, generated by a neural network using the learned scheduling variables, is guaranteed to be stable through a Schur-based parameterization. The architecture combines an encoder for initial state estimation with a state-space representer network that constructs the full set of scheduling-dependent system matrices. For training the NN-SS, we develop a framework that integrates multi-step prediction losses with a state-consistency regularization term, ensuring robustness against drift and improving long-horizon prediction accuracy. The proposed NN-SS is evaluated on benchmark nonlinear systems, and the results demonstrate that the model consistently matches or surpasses classical subspace identification methods and recent gradient-based approaches. These findings highlight the potential of stability-constrained neural LPV identification as a scalable and reliable framework for modeling complex nonlinear systems.
\end{abstract}


\section{Introduction}

System identification (SysID) plays a central role in control engineering, enabling the construction of mathematical models of dynamical systems directly from measured input–output data~\cite{ljung1999system}. Such models are fundamental for prediction, analysis, and controller design across a broad range of applications, including robotics, process automation, and energy systems. Classical SysID approaches have achieved significant success in linear settings, but nonlinear and time-varying processes remain particularly challenging \cite{msaddi,saini2025nonlinear}.

Different types of approaches are observed~\cite{guven2025fuzzy,coelho2024enhancing,node_sysid,dynogp}. A widely adopted framework for handling nonlinearities in a structured manner is the \textit{Linear Parameter-Varying} (LPV) paradigm~\cite{toth2010modeling}. LPV models extend linear system representations by allowing system matrices to depend on scheduling variables, which are often measurable or computable from the operating conditions. By embedding nonlinear dynamics into parameter-dependent state-space equations~\cite{toth2010modeling,sime}, LPV models provide a balance between interpretability and flexibility, making linear control methods applicable to a wider class of systems. However, their performance relies heavily on the availability and quality of scheduling signals; when these are unavailable, noisy, or poorly chosen, both accuracy and model robustness deteriorate~\cite{toth2010modeling}.

In parallel, \textit{neural network-based models} have emerged as data-driven alternatives for SysID~\cite{pillonetto2025deep,narendra1990identification,ferah,cont_sysid,feedforward_sysid,rnn_sysid,lstm_sysid,dai2024deep}. By parameterizing state-transition and output maps with neural networks, these models can capture highly nonlinear behaviors directly from raw data without requiring explicit scheduling variables. Despite their expressive power, such models often lack guarantees of stability~\cite{sime}, which may lead to drift, error accumulation, and reduced reliability in long-horizon simulations, particularly in safety-critical applications. Our earlier work sought to bridge the gap between LPV and neural approaches by introducing hybrid neural state-space formulations~\cite{sertbas2024}. While these models improved flexibility and accuracy, they still depended on predefined scheduling variables and did not fully resolve the stability and robustness issues. Moreover, ensuring accurate long-term predictions under noise and disturbances remains an open challenge.

In this study, we propose a \textit{stable-by-design LPV neural network-based state-space} (NN-SS) model that obviates the need for prior knowledge of scheduling variables by learning both latent states and internal scheduling parameters directly from data. Ensuring stability is a fundamental requirement in state-space modeling, and our framework achieves this through a \textit{stable-by-design} construction of the state-transition matrix. Stability is enforced via a Schur-based parameterization of the state matrix, while the training procedure combines multi-step prediction losses with encoder-based state estimation and state consistency regularization. This approach enhances robustness against drift and improves long-horizon prediction accuracy. The proposed NN-SS is evaluated on benchmark datasets of varying complexity, including a nonlinear two-tank system, a robotic arm, and a multivariable power plant~\cite{daisy}. Results demonstrate that NN-SS consistently matches or surpasses classical subspace identification methods~\cite{van1994n4sid, ljung1999system} and recent gradient-based approaches~\cite{simba}, while producing stable and interpretable latent state trajectories. These findings underscore the potential of stability-constrained neural LPV identification as a reliable and scalable solution for complex nonlinear systems.

\section{The NN-Based LPV SS Model}

The inference procedure of the proposed NN-SS model over $K$ time steps is summarized in Algorithm~\ref{alg:inference}. At its core, the model relies on a discrete-time LPV state-space representation, which enables the system dynamics to adapt to time-varying scheduling parameters:
\begin{equation}
\begin{aligned} \label{eq:lpv_statespacemodel}
\hat{x}(k+1) &= \hat{A}(\rho_k)\,\hat{x}(k) + \hat{B}(\rho_k)\,u(k), \\
\hat{y}(k)   &= \hat{C}(\rho_k)\,\hat{x}(k), \quad \hat{x}(0)=\hat{x}_0
\end{aligned}
\end{equation}
Here, $\hat{x}(k) \in \mathbb{R}^n$ denotes the estimated state vector, $\hat{y}(k) \in \mathbb{R}^m$ is the predicted system output, and $u(k) \in \mathbb{R}^r$ is the system input at time step $k$. 

In this work, the scheduling variable is defined as $\rho_k = \hat{x}(k)$, enabling internal scheduling without the need for explicitly measured scheduling signals. To capture the inherent nonlinearities of the system, two neural networks are embedded into the formulation in \eqref{eq:lpv_statespacemodel}:
\begin{itemize}
    \item \textbf{Encoder NN} ($g: y(0) \mapsto \hat{x}(0)$): Estimates the initial latent state $\hat{x}(0)$ from the system’s first measured output $y(0)$, following the approach in SIME~\cite{sime}. This initialization provides the starting point for state evolution.
    \item \textbf{SS Generator NN} $(f: \rho_k \mapsto (\hat{A}, \hat{B}, \hat{C}))$: Generates the system matrices of the scheduling-dependent state-space model, namely the state-transition matrix $\hat{A}(\rho_k)$, the input-to-state mapping $\hat{B} \in \mathbb{R}^{n \times r}$, and the state-to-output mapping $\hat{C} \in \mathbb{R}^{m \times n}$. To obtain a \emph{stable-by-design} state-transition matrix, we do not generate $\hat{A}$ directly. Instead, the network outputs $\hat{W} \in \mathbb{R}^{2n \times 2n}$ and $\hat{V} \in \mathbb{R}^{n \times n}$, which are combined via a Schur parametrization to construct $\hat{A}(\rho_k)$. The scheduling variable $\rho_k \in \mathbb{R}^{n_\rho}$ enables these matrices to adapt to changing operating conditions, thereby capturing nonlinear and time-varying dynamics.   
\end{itemize}

To summarize, as outlined in Algorithm~\ref{alg:inference}, the initial state $\hat{x}(0)$ is obtained from the encoder network $g(\cdot)$, while at each time step $k$ the state-space generator network $f(\cdot)$ produces the matrices $\hat{A}(\rho_k)$, $\hat{B}(\rho_k)$, and $\hat{C}(\rho_k)$. These matrices are combined with the input sequence $u(k)$ to recursively update the state and predict the output, in accordance with the representation in \eqref{eq:lpv_statespacemodel}. The subsequent subsections present the architectures and design considerations of the employed neural networks in detail.

\begin{algorithm}[b]
\caption{Inference procedure for NN-SS model}
\label{alg:inference}
\begin{algorithmic}[1]
\STATE \textbf{Input:} Input sequence $\{u(k)\}_{k=0}^K$, output $y(0)$
\STATE Initialize: $\hat{x}(0) \gets g(y(0))$
\FOR{$k = 0$ to $K$}
    \STATE $\rho_k \gets \hat{x}(k)$ \COMMENT{internal scheduling}
    \STATE $[\hat{A}(\rho_k), \hat{B}(\rho_k), \hat{C}(\rho_k)] \gets f(\rho_k)$
    \STATE $\hat{x}(k+1) \gets \hat{A}(\rho_k)\hat{x}(k) + \hat{B}(\rho_k)u(k)$
    \STATE $\hat{y}(k) \gets \hat{C}(\rho_k)\hat{x}(k)$
\ENDFOR
\STATE \textbf{Output:} Predicted trajectory $\{\hat{y}(k)\}_{k=0}^K$
\end{algorithmic}
\end{algorithm}

\subsection{The Encoder Neural Network}
The encoder network $g(.)$ follows a standard multilayer perceptron (MLP) that can be adapted to the characteristics of the dataset. It consists of fully connected layers with weights and biases, but no activation functions between layers, since the goal is to obtain a linear mapping from the outputs to the latent states. As an example, its first layer is defined as:
\begin{equation} \label{eq:encoder_first}
h_1 = w_1 \cdot y(k) + b_1,
\end{equation}
where $w_1 \in \mathbb{R}^{\mathcal{N}_1 \times m}$ is the weight matrix, $\mathcal{N}_1$ is the number of neurons in the first hidden layer, and $b_1 \in \mathbb{R}^{\mathcal{N}_1 \times 1}$ is the bias vector. The final encoder output is written as:
\begin{equation} \label{eq:encoder_output}
\hat{x}(k) = w_2 \cdot h_1 + b_2,
\end{equation}
where $w_2 \in \mathbb{R}^{n \times \mathcal{N}_1}$ and $b_2 \in \mathbb{R}^{n \times 1}$ yield the estimated states $\hat{x}(k) \in \mathbb{R}^{n \times m}$.

\subsection{The State-Space Generator Neural Network}
Ensuring stability is a fundamental requirement in state-space modeling. In our framework, this is achieved through a \textbf{stable-by-design} construction of the state-transition matrix. For systems of the form~\eqref{eq:lpv_statespacemodel}, stability requires that \(A(\rho_k)\) be Schur, i.e.,
\begin{equation}
|\lambda_i\big(A(\rho_k)\big)| < 1, \quad \forall i = 1, \dots, n.
\end{equation}
To satisfy this property, we build upon the free parametrization of Schur matrices like in SIMBa~\cite{simba,simbaext}. Instead of generating \(A\) directly, our model produces auxiliary matrices \(W\) and \(V\), which are then mapped into a Schur-stable \(A\) by the following result.

\medskip
\noindent \textbf{Schur parametrization:}  
Let \(W \in \mathbb{R}^{2n \times 2n}\), \(V \in \mathbb{R}^{n \times n}\), \(\tilde{\epsilon} \in \mathbb{R}\), and \(0 < \gamma \leq 1\). Define
\begin{equation} \label{eq:S_def}
S := W^\top W + \epsilon I_{2n}, \quad \epsilon = \exp(\tilde{\epsilon}).
\end{equation}
Then,
\begin{equation} \label{eq:schur}
A = S_{12} 
\left[ 
\frac{1}{\gamma^2}
\begin{pmatrix}
S_{11} & 0 \\
0 & S_{22}
\end{pmatrix}
+ V - V^\top 
\right]^{-1}
\end{equation}
is Schur stable, with all eigenvalues strictly inside the disk of radius \(\gamma\). A detailed derivation is given in~\cite{simbaext,MilCal:13}.

\medskip
\noindent
\textbf{Implementation in our model:}  
The SS generator NN $f(.)$ employs a similar MLP structure but includes activation functions to exploit the full expressive power of NNs. As an example, its first hidden layer is given by:
\begin{equation} \label{eq:firstSSNN}
h_1 = \phi_1(w_1 \cdot \rho_k + b_1),
\end{equation}
where $w_1 \in \mathbb{R}^{\mathcal{N}_1 \times n_\rho}$ and $b_1 \in \mathbb{R}^{\mathcal{N}_1 \times 1}$ are the weight and bias, and $\phi_1$ is the activation function. The final output layer is defined as:
\begin{equation} \label{eq:fourthSSNN}
output = w_2 \cdot h_1 + b_2,
\end{equation}
with $w_2 \in \mathbb{R}^{\mathcal{N}_2 \times \mathcal{N}_1}$ and $b_2 \in \mathbb{R}^{\mathcal{N}_2 \times 1}$ representing the layer weights and biases. No activation is applied at this stage to avoid restricting the prediction range. The neuron count $\mathcal{N}_2$ equals the dimensionality of all required parameters so that the $output \in \mathbb{R}^{5n^2 + nr + mn}$ contains all model components. 

The state-space generator NN outputs a vector that is partitioned as
\[
output = \bigl[\hat{W},\, \hat{V},\, \hat{B},\, \hat{C}\bigr],
\]
where the first \(4n^2\) entries define \(\hat{W}\), the next \(n^2\) entries define \(\hat{V}\), and the remaining terms correspond to \(\hat{B}\) and \(\hat{C}\). After reshaping into matrices, \(\hat{W} \in \mathbb{R}^{2n \times 2n}\) and \(\hat{V} \in \mathbb{R}^{n \times n}\) are substituted into~\eqref{eq:schur} to obtain the Schur-stable state-transition matrix \(\hat{A} \in \mathbb{R}^{n \times n}\). The subsequent segments of the output vector are reshaped into \[
\hat{B} \in \mathbb{R}^{n \times r}, 
\quad
\hat{C} \in \mathbb{R}^{m \times n},
\] which provide the input-to-state and state-to-output mappings, respectively. Together, \((\hat{A}, \hat{B}, \hat{C})\) yield the complete discrete-time LPV state-space representation of~\eqref{eq:lpv_statespacemodel}.

\section{NN-SS Training Strategy}

The proposed training procedure for the NN-SS, including mini-batch processing and parameter updates, is summarized in Algorithm~\ref{alg:ssnn}\footnotemark{}.
In the following subsections, we provide the details of the training dataset construction and the design of the loss function used to optimize the network parameters.

\footnotetext{MATLAB implementation. [Online]. Available: \url{https://github.com/AbiyeshanTR/NNSS}}

% Implementation details can be found in the \href{https://github.com/AbiyeshanTR/SSNN-Library-A.-E.-Sertbas-T.-Kumbasar-IWW-2025-}{\textbf{GitHub repository}}.

\begin{algorithm}[t]
\caption{Training procedure for NN-SS model}
\label{alg:ssnn}
\begin{algorithmic}[1]
\STATE \textbf{Input:} $\mathcal{D}_{\text{train}}$ (trajectories $\{\mathbf{T}_k\}$, length $L$, stride $s$), order $n$, optimizer ADAM

\FOR{each epoch}
  \FOR{each mini-batch $Z \subset \mathcal{D}_{\text{train}}$}
  \STATE Initialize: $\hat{x}^{(t)}(0)\gets g\!\left(y^{(t)}(0)\right),\ \forall t\in Z$
    \FOR{$k = 0$ to $L$, each trajectory $t \in Z$}
    
      \STATE $\rho_k^{(t)} \gets \hat{x}^{(t)}(k)$ \COMMENT{internal scheduling}
      \STATE $[\hat{A}(\rho_k),\hat{B}(\rho_k),\hat{C}(\rho_k)]\!\gets\!f(\rho_k^{(t)})$
      \STATE $\hat{x}^{(t)}(k+1) \gets \hat{A}(\rho_k)\,\hat{x}^{(t)}(k) + \hat{B}(\rho_k)\,u^{(t)}(k)$
      \STATE $\hat{y}^{(t)}(k) \gets \hat{C}(\rho_k)\,\hat{x}^{(t)}(k)$
      \STATE $\hat{x}^{(t,\mathrm{ENC})}(k) \gets g\!\left(y^{(t)}(k)\right)$
    \ENDFOR
    \STATE Compute total loss $L_{\text{total}} = L_{\text{resp}} + \lambda \cdot\,L_{\text{state}}$
    \STATE Backpropagate $L_{\text{total}}$, update parameters via ADAM
  \ENDFOR
  \STATE Validate; early stop on validation RMSE, shuffle $\mathcal{D}_{\text{train}}$
\ENDFOR
\STATE \textbf{Output:} Trained $f$, $g$
\end{algorithmic}
\end{algorithm}

\subsection{Data Preparation}
Capturing multi-step system dynamics necessitates preparing a dataset that includes sequences of consecutive input–output measurements. This enables the NN-SS model to learn the underlying state-space relationships effectively. To this end, we start from a raw dataset that records all measured outputs and inputs over time. Formally, we assume having a raw dataset
\(\mathcal{D} = \{\mathbf{X}_k\}_{k=1}^K\), where
\begin{equation}
\mathbf{X}_k = \big(y_1(k), \ldots, y_m(k), \, u_1(k), \ldots, u_r(k)\big),
\end{equation}
each row \(\mathbf{X}_k\) concatenates the current system output \(y_k \in \mathbb{R}^m\) and input \(u_k \in \mathbb{R}^r\).

To construct the training dataset, overlapping trajectories of length \(L\) are extracted using a sliding window with stride \(s\), following the approach in SIMBa~\cite{simba}:
\begin{equation}
\begin{aligned}
\mathcal{D}_{\text{train}} &= \{\mathbf{T}_k\}_{k=1}^N, 
\quad \mathbf{T}_k = (\mathbf{X}_k, \mathbf{X}_{k+1}, \ldots, \mathbf{X}_{k+L-1}), \\
k &= 1,\, 1+s,\, 1+2s,\, \ldots,\, 1 + \Big\lfloor \frac{K-L}{s} \Big\rfloor s,
\end{aligned}
\end{equation}
where each trajectory \(\mathbf{T}_k\) consists of \(L\) consecutive samples, and \(N\) denotes the total number of trajectories. The resulting training set \(\mathcal{D}_{\text{train}}\) is shuffled and partitioned into mini-batches \(Z\) of the desired size. 

\subsection{Loss Function}

The training objective of the proposed NN-SS model is formulated as a composite loss function that combines multi-step output prediction accuracy with latent state consistency:

\begin{equation}
L_{\text{total}} = L_{\text{response}} + \lambda \cdot L_{\text{state}},
\end{equation}
where $\lambda$ is a regularization coefficient balancing the contribution of the auxiliary state loss $L_{\text{state}}$. 

The primary term, $L_{\text{response}}$, penalizes the multi-step prediction error over each trajectory in a mini-batch $Z$, following the approach in SIMBa~\cite{simba, simbaext}:
\begin{equation}
\label{eq:response_loss}
L_{\text{response}} = \frac{1}{|Z|} \sum_{t \in Z} 
\frac{1}{L \cdot m} \sum_{k=0}^{L} \sum_{i=1}^{m}
\mathcal{L}_{\text{2}}\!\left( y^{(t)}_{k,i}, \hat{y}^{(t)}_{k,i} \right),
\end{equation}
where $\mathcal{L}_{2} $ is the L2 loss, k indexes the time step within the trajectory, $i$ indexes the output channel for MIMO systems, and $t$ identifies the trajectory in the mini-batch.

To encourage consistency and stability of the latent state evolution, an auxiliary regularization term $L_{\text{state}}$ is introduced, inspired by SIME~\cite{sime}. This term penalizes discrepancies between the propagated states and those inferred by the encoder, excluding the initial state:
\begin{equation}
\label{eq:state_loss}
L_{\text{state}} = \frac{1}{|Z|} \sum_{t \in Z} 
\frac{1}{(L-1)\cdot n} \sum_{k=1}^{L} \sum_{j=1}^{n} 
\mathcal{L}_{\text{2}}\!\left( \hat{x}^{(t)}_{k,j}, \hat{x}^{(t,\mathrm{ENC})}_{k,j} \right),
\end{equation}
where $\hat{x}^{(t)}_k$ denotes the state propagated through the model dynamics and $\hat{x}^{(t,\mathrm{ENC})}_k$ the encoder-derived estimate at the same step. This regularization enforces consistency between the two, reducing drift and improving robustness over long horizons and noisy data.

\section{Comparative System Identification Results}

The proposed NN-SS model is benchmarked against three classical subspace identification methods---N4SID~\cite{van1994n4sid}, SSEST, and SSREGEST~\cite{ljung1999system}---from MATLAB’s System Identification Toolbox, and the gradient-based SIMBa~\cite{simba, simbaext}. Datasets are normalized with z-score scaling, and NN-SS  hyperparameters (layers, widths, activations) are tuned via Bayesian optimization; in this work, sigmoid activation is employed in $f(\cdot)$. SIMBa is initialized from N4SID, with the state-transition matrix $A$ obtained by optimizing $W$ and $V$ via Frobenius norm minimization~\cite{MilCal:13} using MATLAB’s \texttt{fmincon}. To ensure fair comparison, model order is set to the N4SID estimate, with $\pm 1$ order variations reported to assess generalization. Both NN-SS and SIMBa use early stopping on validation RMSE to prevent overfitting.  

Each model is trained with 10 different random seeds to provide robust performance estimates. Model accuracy is quantified using the root mean square error (RMSE) averaged across output channels, providing a single comparison metric across systems.

\subsection{SysID Performance Analysis of the Two-Tank System} 
The two-tank system from MATLAB’s System Identification Toolbox exhibits nonlinear, coupled flow dynamics with a sampling time $T_s$ of 0.2 s. The dataset was segmented into overlapping trajectories of 80 samples with a stride of 2 to increase training diversity. Training followed the configuration in Table~\ref{tab:training_config_twotank}.

\begin{table}[b]
\caption{Training Configuration for Two-Tank System}
\label{tab:training_config_twotank}
\centering
\resizebox{\columnwidth}{!}{%
\begin{tabular}{|l|l|l|l|}
\hline
Parameter & Value & Parameter & Value \\
\hline
Number of states ($n$) & 2, 3 (N4SID), 4 & Batch size ($Z$) & 64 \\
Number of epochs & 200 & Learning rate & 0.001 \\
Total trajectories ($N$) & 711 & Regularization ($\lambda$) & 0.01 \\
$f(\cdot)$ architecture & 2 hidden layers & $g(\cdot)$ architecture & 2 hidden layers \\
\hline
\end{tabular}%
}
\end{table}

\begin{figure}[t]
\centering
\includegraphics[width=\columnwidth]{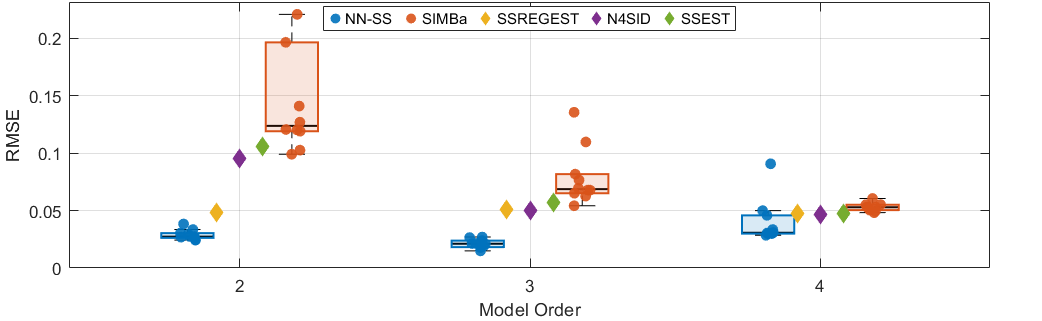}
\caption{Test RMSE for different methods and training random seeds across model orders on the two-tank dataset.}
\label{NMSE twotank}
\end{figure}

Fig.~\ref{NMSE twotank} reports the test RMSE across model orders and random seeds. The proposed NN-SS consistently outperforms SIMBa and classical subspace methods, particularly at $n=2$ and $n=3$, where the others show higher variance and larger errors. NN-SS delivers the best median accuracy and greater robustness across seeds, reflecting the advantages of stability constraints and trajectory-based training.

\begin{figure}[b]

\centering
\includegraphics[width=\columnwidth]{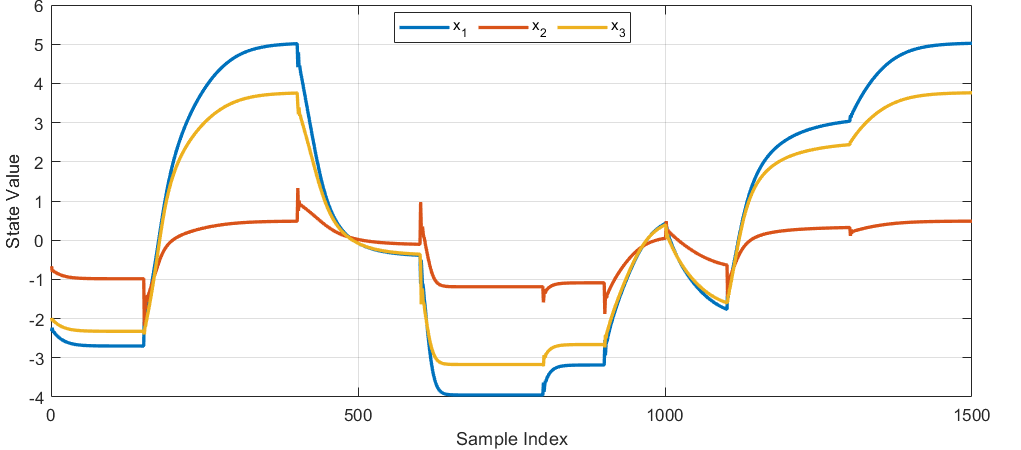}
\caption{Estimated latent states for the best NN-SS run on the two-tank dataset.}
\label{twotank state}
\end{figure}

Table~\ref{tab:twotank_all} reports the RMSE-wise best-performing seeds for NN-SS and SIMBa for model order \(n=3\). Fig.~\ref{twotank state} shows latent states from the best-performing NN-SS run. Two states closely follow the measured outputs, indicating the model has effectively captured and encoded the dominant dynamics in latent space while preserving stability.

\begin{figure}[t]
\centering
\includegraphics[width=\columnwidth]{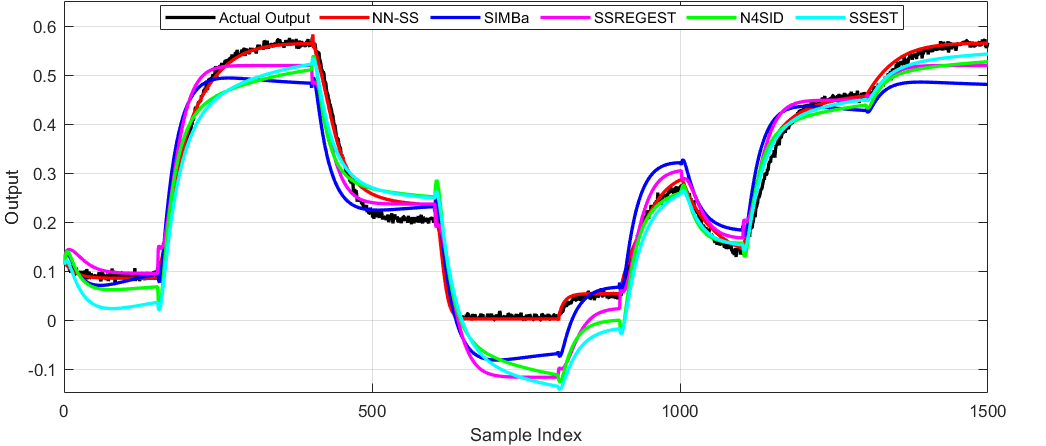}
\caption{Simulation mode for best model seeds in two-tank system.}
\label{twotank sim}
\end{figure}

Fig.~\ref{twotank sim} compares long-horizon simulations for the best model from each method. While all models remain stable and follow the measured output, NN-SS tracks the signal most closely; all models exhibit a small undershoot, most noticeable during the initial rise. This likely reflects a mismatch between the learned dynamics and the plant’s fast transient, amplified by data-driven training and a multi-step loss that prioritizes long-horizon accuracy over short-term fidelity. SIMBa’s weaker results relative to N4SID suggest that the chosen \(A\) matrix initialization and training on randomly sampled trajectories can be crucial in the training of SIMBa.

In summary, the NN-SS not only achieves the lowest RMSE in both validation and test sets but also produces state trajectories that align well with the physical system. Its stable-by-design architecture ensures bounded trajectories in simulation mode, while classical methods, although competitive in short-term fits, underperformed over long horizons.

\begin{table}[b]
\centering
\caption{RMSE Comparison: Two Tank System}
\label{tab:twotank_all}
\setlength{\tabcolsep}{14pt}
\renewcommand{\arraystretch}{0.8}
\footnotesize
\begin{tabular*}{\columnwidth}{|@{\extracolsep{\fill}}|l|c|c|c|}
\hline
\textbf{Model} & \textbf{Train} & \textbf{Val} & \textbf{Test} \\
\hline
NN-SS     & \textbf{0.02} & \textbf{0.01} & \textbf{0.01} \\
SIMBa     & 0.05 & 0.06 & 0.05 \\
SSREGEST  & 0.04 & 0.04 & 0.05 \\
N4SID     & 0.04 & 0.05 & 0.05 \\
SSEST     & 0.04 & 0.05 & 0.06 \\
\hline
\end{tabular*}
\end{table}

\subsection{SysID Performance Analysis of the Robot Arm} 
The robot arm system, provided in MATLAB's System Identification Toolbox and characterized by high-dimensional, nonlinear, and coupled joint dynamics, serves as a challenging benchmark for black-box modeling. To suppress high-frequency noise and reduce redundancy, the dataset was downsampled by a factor of 10, with $T_s$ of 0.005 s. It was then segmented into overlapping trajectories of 100 samples with a stride of 5, ensuring diverse yet temporally coherent training sequences. This supports robust long-horizon multi-step learning and aids in lowering computational cost. Training was performed using the configuration in Table~\ref{tab:training_config_robot arm}.

\begin{figure}[b] \centering \includegraphics[width=\columnwidth]{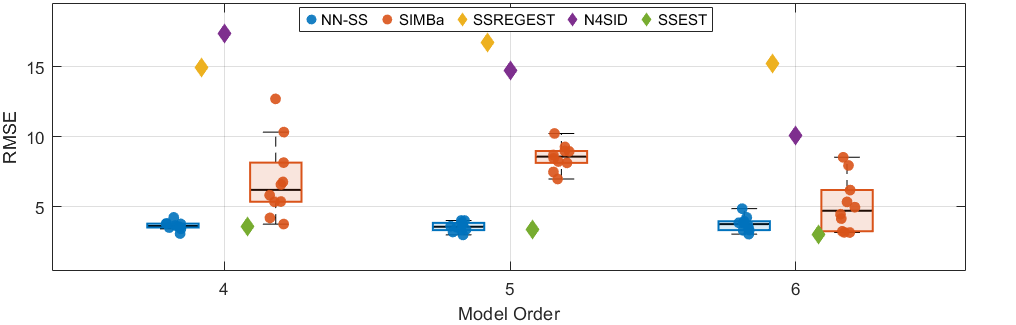} \caption{Test RMSE for different methods and training random seeds across model orders on the robot arm dataset.}  \label{NMSE robotarm} \end{figure}

After the training phase, Fig.~\ref{NMSE robotarm} shows the RMSE test across model
orders and seeds. In most configurations, both the NN-SS and SSEST achieve performance that is competitive with each other and superior when compared to both SIMBa and classical subspace methods.

\begin{figure}[t] \centering \includegraphics[width=\columnwidth]{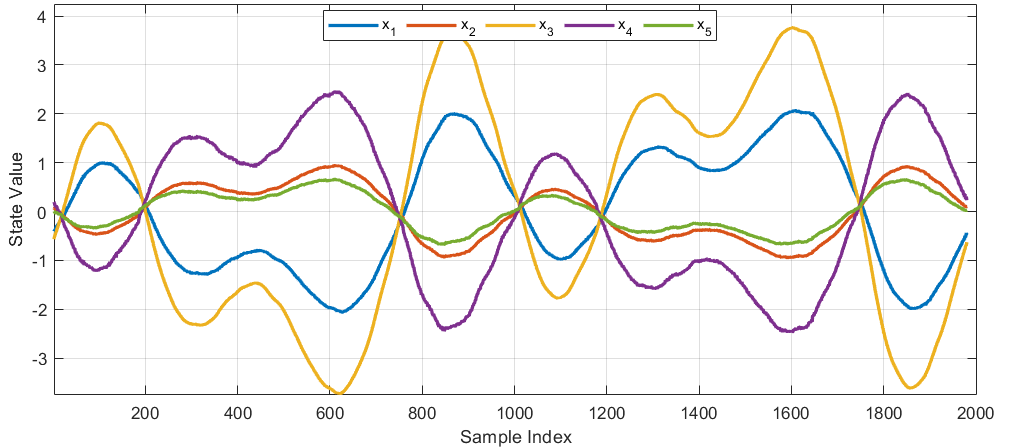} \caption{Evaluation of states for best NN-SS seed in robot arm.} \label{robot state} \end{figure}

Table~\ref{tab:robot_all} reports the RMSE-wise best-performing seeds for NN-SS and SIMBa at model order \(n=5\). Fig.~\ref{robot state} shows the NN-SS’s latent states for the robot arm system. The states remain bounded and smooth, with distinct roles: a dominant low-frequency state follows slow trends, oscillatory states sharpen transients, and small-amplitude states act as fast residual correctors.

\begin{table}[b]
\caption{Training Configuration for Robot Arm System}
\label{tab:training_config_robot arm}
\centering
\resizebox{\columnwidth}{!}{%
\begin{tabular}{|l|l|l|l|}
\hline
Parameter & Value & Parameter & Value \\
\hline
Number of states ($n$) & 4, 5 (N4SID), 6 & Batch size ($Z$) & 32 \\
Number of epochs & 200 & Learning rate & 0.001 \\
Total trajectories ($N$) & 377 & Regularization ($\lambda$) & 0.01 \\
$f(\cdot)$ architecture & 4 hidden layers & $g(\cdot)$ architecture & 4 hidden layers \\
\hline
\end{tabular}%
}
\end{table}

\begin{table}[t]
\centering
\caption{RMSE Comparison: Robot Arm}
\label{tab:robot_all}
\setlength{\tabcolsep}{14pt}
\renewcommand{\arraystretch}{0.9}
\footnotesize
\begin{tabular*}{\columnwidth}{|@{\extracolsep{\fill}}l|c|c|c|}
\hline
\textbf{Model} & \textbf{Train} & \textbf{Val} & \textbf{Test} \\
\hline
NN-SS     & 5.26 & 5.16 & \textbf{3.00} \\
SIMBa     & 5.05 & 4.99 & 6.99 \\
SSREGEST  & 14.93 & 14.65 & 16.71 \\
N4SID     & 10.58 & 10.36 & 14.72 \\
SSEST     & \textbf{2.76} & \textbf{2.66} & 3.38 \\
\hline
\end{tabular*}
\end{table}

\begin{figure}[b] \centering \includegraphics[width=\columnwidth]{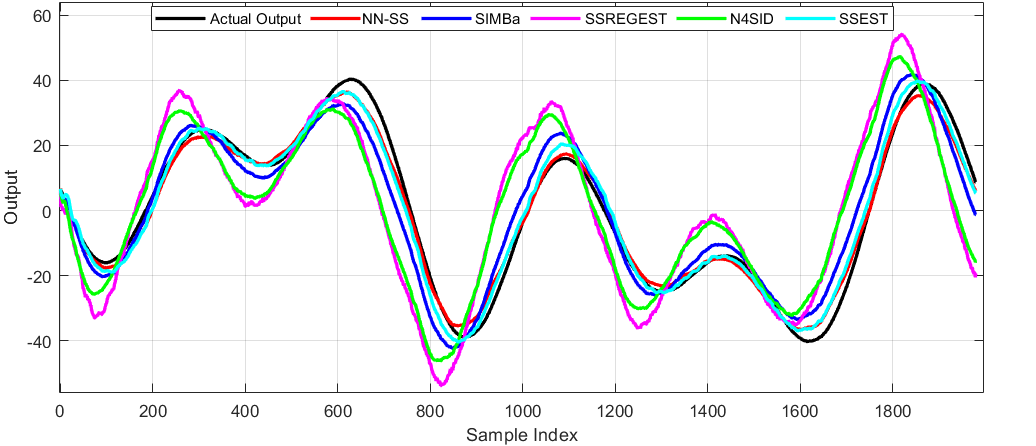} \caption{Simulation mode for best model seeds in robot arm system.} \label{robot sim} \end{figure}

Fig. \ref{robot sim} compares the simulation mode performance of all models. The NN-SS and SSEST trace the measured output most closely over the horizon—especially at the peaks and troughs—resulting in the lowest test RMSEs. SIMBa tends to generalize the dynamics better than subspace methods, while N4SID/SSREGEST shows phase lag and amplitude bias.

In summary, compared to the earlier two-tank results, SIMBa outperforms N4SID on this dataset, suggesting that initialization and the use of randomly sampled training trajectories can significantly benefit SIMBa. SSEST’s performance further supports this, as an N4SID–based method with additional optimization, achieving more precise predictions and lower errors. The NN-SS, however, consistently matches or surpasses all methods, highlighting its robustness and adaptability across different system dynamics.

\subsection{SysID Performance Analysis of the Power Plant} 
The simulation study uses the Power Plant dataset from KU Leuven’s DaISy~\cite{daisy} repository, featuring slow and coupled multivariable dynamics with $T_s$ of 1228.8 s. As in SIMBa~\cite{simba}, no trajectory segmentation was applied; the full dataset was reused during training to preserve temporal continuity and match SIMBa's original modeling assumptions. Training
was performed using the configuration in Table \ref{tab:training_config_powerplant} for learning the dynamics of the nonlinear power plant.

\begin{table}[t]
\caption{Training Configuration for Power Plant System}
\label{tab:training_config_powerplant}
\centering
\resizebox{\columnwidth}{!}{%
\begin{tabular}{|l|l|l|l|}
\hline
Parameter & Value & Parameter & Value \\
\hline
Number of states ($n$) & 3, 4 (N4SID), 5 & Batch size ($Z$) & 1 \\
Number of epochs & 1000 & Learning rate & 0.001 \\
Total trajectories ($N$) & 1 & Regularization ($\lambda$) & 0.01 \\
$f(\cdot)$ architecture & 5 hidden layers & $g(\cdot)$ architecture & 4 hidden layers \\
\hline
\end{tabular}%
}
\end{table}

\begin{figure}[t] \centering \includegraphics[width=\columnwidth]{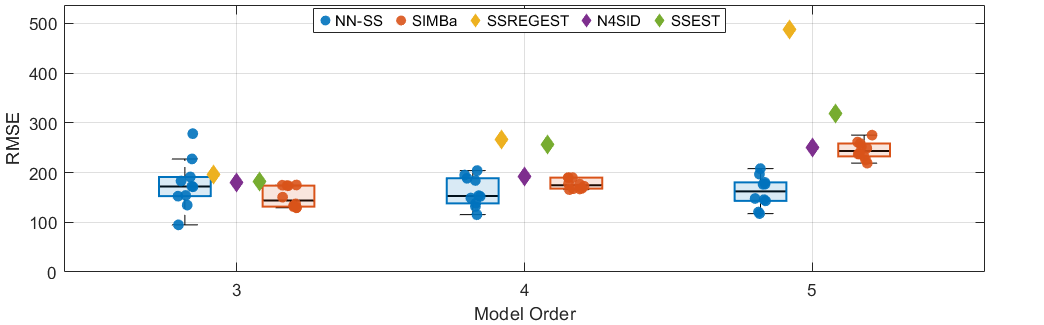} \caption{Test RMSE for different methods and training random seeds across model orders on the power plant dataset.}  \label{NMSE powerplant} \end{figure}

Fig.~\ref{NMSE powerplant} shows the test RMSE across model orders and seeds. The NN-SS consistently performs with the lowest error, exhibiting deviations at lower orders, while maintaining stable performance across seeds. SIMBa demonstrates competitive accuracy in certain configurations, whereas classical methods exhibit higher median errors, highlighting the benefits of stability constraints and internal scheduling in NN-SS.

\begin{figure}[b] \centering \includegraphics[width=\columnwidth]{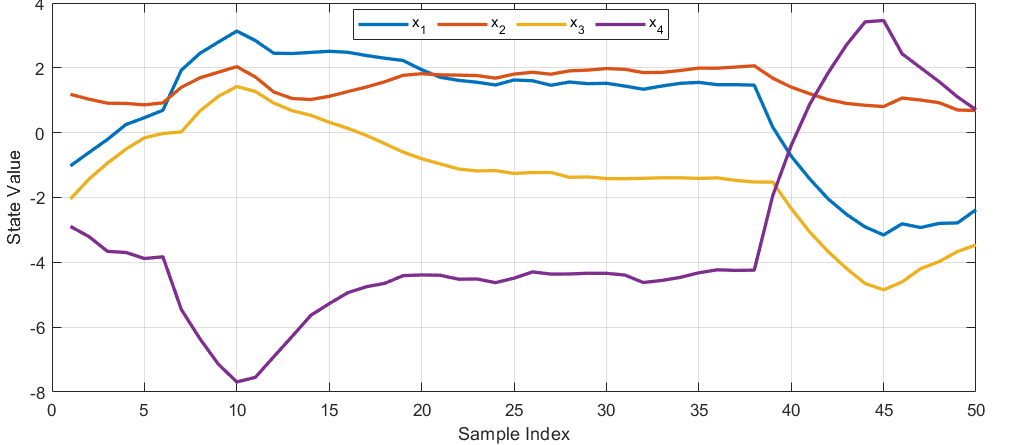} \caption{Evaluation of states for best NN-SS seed in power plant.} \label{powerplant state} \end{figure}

Table~\ref{tab:powerplant_all} reports the best RMSE seeds for NN-SS and SIMBa at model order \(n=4\). Fig.~\ref{powerplant state} shows that the slowly varying states $x_1, x_2$ track the gradual trends in $y_2$ and $y_3$, while $x_3$ provides a small offset correction. During the main transient ($k\!\approx\!38$–46), the states evolve in a coordinated fashion: $x_4$ makes a sharp excursion that anticipates the fast rise/decay in the outputs, while $x_1, x_2$ reverse slope more gradually, capturing the slower return to steady conditions. 
Turning points in the states align with peaks and troughs of the outputs, showing that NN-SS has learned a compact latent representation separating fast and slow modes. The states remain bounded and drift-free, indicating well-posed dynamics and effective initial-state mapping.

\begin{figure}[t] \centering \includegraphics[width=\columnwidth]{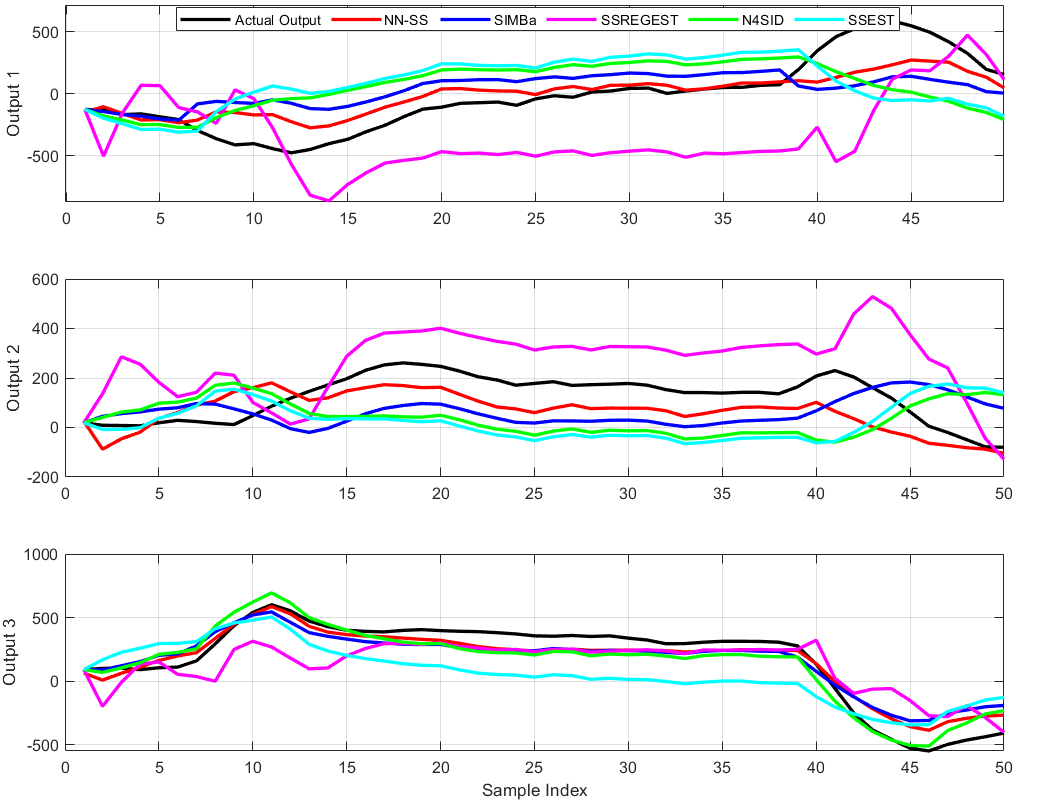} \caption{Simulation mode for best model seeds in power plant system.} \label{powerplant sim} \end{figure}

Fig. \ref{powerplant sim} compares the simulation mode performance of all models. Across the three outputs, NN-SS and SIMBa provide the closest overall matches to the measured signals. The linear baselines N4SID/SSEST capture slow trends but exhibit phase lag and amplitude bias on fast segments, while SSREGEST shows the largest deviations.  More specifically, we observe that: 
\begin{itemize}
    \item \textbf{\emph{Output~1:}}  
    All models capture the slow drift up to $k \!\approx\! 40$, but SSREGEST develops a strong negative bias near $k \!\approx\! 10$, leading to persistently high RMSE. N4SID and SSEST are smoother but miss major transients, showing clear lag. NN-SS tracks both the timing and magnitude of the rise–fall sequence most closely, though with minor dropouts. SIMBa comes close to NN-SS but suffers from a systematic gain mismatch.
    
    \item \textbf{\emph{Output~2:}}  
    During the early transient ($k \!\approx\! 1$–$20$), NN-SS and SIMBa remain close to the measured signal, while N4SID and SSEST consistently underestimate the level. In the quasi–steady-state region, all models show gain errors. Around the late transient ($k \!\approx\! 40$–$50$), SSREGEST overshoots heavily before collapsing, which dominates its RMSE. NN-SS captures both onset and decay of the transient most accurately, while other methods retain a negative bias.
    
    \item \textbf{\emph{Output~3:}}  
    While all models reproduce the general dynamics, gain errors persist throughout. N4SID, SSEST, and SSREGEST miss the sharp drop after $k \!\approx\! 40$, resulting in lag and amplitude mismatch. NN-SS and SIMBa preserve both curvature and timing, yielding the lowest errors, with NN-SS showing slightly stronger alignment during the steepest changes.
\end{itemize}

In summary, NN-SS yields the lowest test RMSE by accurately reproducing both amplitude and timing of transients. SIMBa is competitive but degrades on steep segments and can become trapped in local minima due to N4SID initialization. For SIMBa results without initialization, see \cite{simba}, \cite{simbaext}. The high RMSE across models indicates insufficient input excitation and operating coverage when obtaining data from the plant, leaving key modes weakly excited and causing lag, gain errors, or unstable transients under unseen conditions. Findings highlight the importance of carefully selected input signals and diverse operating scenarios to ensure reliable estimates.

\begin{table}[t]
\centering
\caption{RMSE Comparison: Power Plant}
\label{tab:powerplant_all}
\setlength{\tabcolsep}{12pt}
\renewcommand{\arraystretch}{0.95}
\footnotesize

% --- Channel y1 ---
\begin{tabular*}{\columnwidth}{|@{\extracolsep{\fill}}l|c|c|c|}
\hline
\multicolumn{4}{|c|}{\textbf{Performance for Channel $y_1$}} \\
\hline
\textbf{Model} & \textbf{Train} & \textbf{Val} & \textbf{Test} \\
\hline
NN-SS     & 84.84 & 85.07 & \textbf{188.33} \\
SIMBa    & \textbf{20.85} & \textbf{78.35} & 247.51 \\
SSREGEST & 273.27 & 201.49 & 440.91 \\
N4SID    & 60.43 & 82.14 & 303.66 \\
SSEST    & 49.58 & 105.80 & 345.51 \\
\hline
\end{tabular*}

\vspace{1mm}

% --- Channel y2 ---
\begin{tabular*}{\columnwidth}{|@{\extracolsep{\fill}}l|c|c|c|}
\hline
\multicolumn{4}{|c|}{\textbf{Performance for Channel $y_2$}} \\
\hline
\textbf{Model} & \textbf{Train} & \textbf{Val} & \textbf{Test} \\
\hline
NN-SS      & 50.62 & \textbf{39.11} & \textbf{110.00} \\
SIMBa    & 25.40 & 84.87 & 140.76 \\
SSREGEST & 76.58 & 69.92 & 174.10 \\
N4SID    & 24.22 & 94.07 & 167.93 \\
SSEST    & \textbf{19.17} & 88.74 & 178.94 \\
\hline
\end{tabular*}

\vspace{1mm}

% --- Channel y3 ---
\begin{tabular*}{\columnwidth}{|@{\extracolsep{\fill}}l|c|c|c|}
\hline
\multicolumn{4}{|c|}{\textbf{Performance for Channel $y_3$}} \\
\hline
\textbf{Model} & \textbf{Train} & \textbf{Val} & \textbf{Test} \\
\hline
NN-SS     & 34.19 & \textbf{55.93} & \textbf{67.90} \\
SIMBa    & 37.70 & 67.53 & 104.21 \\
SSREGEST & 59.15 & 108.02 & 187.69 \\
N4SID    & 27.83 & 118.37 & 105.15 \\
SSEST    & \textbf{16.23} & 119.13 & 246.04 \\
\hline
\end{tabular*}

\end{table}

\section{Conclusion and Future Work}

We proposed a stable-by-design NN-SS model that simultaneously learns latent states and LPV dynamics using multi-step simulation losses and internal scheduling, enabling efficient system identification. Across benchmark studies—including a two-tank system, a robotic arm, and a multivariable power plant—the NN-SS consistently achieved the lowest or competitive RMSE while producing stable and interpretable state trajectories. Gradient-based approaches, such as SIMBa, performed well with careful initialization, whereas classical subspace methods captured slow dynamics but often lagged during fast transients.

Future work will focus on analyzing the learned latent dimensions from a control engineering perspective, enabling their use in controller design. Additionally, uncertainty quantification will be investigated to support safe and reliable real-time deployment of NN-SS models.


\begin{thebibliography}{00} 

\bibitem{ljung1999system} L. Ljung, \textit{System Identification: Theory for the User}, 2nd~ed. Upper Saddle River, NJ, USA: Prentice Hall, 1999. 

\bibitem{msaddi}
S. Msaddi and T. Kumbasar, "Expanding conformal prediction to system identification," \textit{Pattern Recognition}, p. 111758, 2025.

\bibitem{saini2025nonlinear} 
K. Saini, N. Kumar, B. Bhushan, and R. Kumar, “Nonlinear complex dynamic system identification based on a novel recurrent neural network,” \textit{Soft Computing}, pp. 1–20, 2025.

\bibitem{guven2025fuzzy} 
Y. Güven and T. Kumbasar, “Fuzzy Logic Strikes Back: Fuzzy ODEs for Dynamic Modeling and Uncertainty Quantification,” \textit{IEEE Transactions on Artificial Intelligence}, vol. 6, pp. 2788 - 2797, 2025.

\bibitem{coelho2024enhancing} 
C. Coelho, M. F. P. Costa, and L. L. Ferrás, “Enhancing continuous time series modelling with a latent ODE-LSTM approach,” \textit{Applied Mathematics and Computation}, vol. 475, p. 128727, 2024.

\bibitem{node_sysid} 
Y. Yang and H. Li, “Neural ordinary differential equations for robust parameter estimation in dynamic systems with physical priors,” \textit{Applied Soft Computing}, vol. 169, p. 112649, 2025.

\bibitem{dynogp} 
A. Benavoli, D. Piga, M. Forgione, and M. Zaffalon, “dynoGP: Deep Gaussian Processes for dynamic system identification,” \textit{arXiv preprint arXiv:2502.05620}, 2025.

\bibitem{toth2010modeling} R. Tóth, \textit{Modeling and Identification of Linear Parameter-Varying Systems}. Lecture Notes in Control and Information Sciences, vol. 403. Springer, 2010. 

\bibitem{sime} Y. Bao, J. Mohammadpour Velni, A. Basina, and M. Shahbakhti, `Identification of state-space linear parameter-varying models using artificial neural networks,'' \textit{IFAC-PapersOnLine}, vol. 53, no. 2, pp. 5286--5291, 2020. 

\bibitem{narendra1990identification} K. S. Narendra and K. Parthasarathy, `Identification and control of dynamical systems using neural networks," in \textit{IEEE Trans. Neural Netw.}, vol. 1, pp. 4--27, 1990. 


\bibitem{ferah}
M. A. Ferah and T. Kumbasar, "Introducing Interval Neural Networks for Uncertainty-Aware System Identification," in \textit{2025 7th International Congress on Human-Computer Interaction, Optimization and Robotic Applications}, 2025.

\bibitem{cont_sysid} 
M. Forgione and D. Piga, “Continuous-time system identification with neural networks: Model structures and fitting criteria,” \textit{European Journal of Control}, vol. 59, pp. 69–81, 2021.

\bibitem{dai2024deep} 
T. Dai, K. Aljanaideh, R. Chen, R. Singh, A. Stothert, and L. Ljung, “Deep Learning of Dynamic Systems using System Identification Toolbox\texttrademark,” \textit{IFAC-PapersOnLine}, vol. 58, no. 15, pp. 580–585, 2024.

\bibitem{feedforward_sysid} 
T. A. Tutunji, “Parametric system identification using neural networks,” \textit{Applied Soft Computing}, vol. 47, pp. 251–261, 2016.


\bibitem{pillonetto2025deep} 
G. Pillonetto, A. Aravkin, D. Gedon, L. Ljung, A. H. Ribeiro, and T. B. Schön, “Deep networks for system identification: a survey,” \textit{Automatica}, vol. 171, p. 111907, 2025.

\bibitem{rnn_sysid} 
M. Forgione, A. Muni, D. Piga, and M. Gallieri, “On the adaptation of recurrent neural networks for system identification,” \textit{Automatica}, vol. 155, p. 111092, 2023.

\bibitem{lstm_sysid} 
Y. Wang, “A new concept using LSTM Neural Networks for dynamic system identification,” in \textit{Proc. 2017 American Control Conference (ACC)}, pp. 5324–5329, 2017.

\bibitem{sertbas2024} A. E. Sertbaş, A. Köklü, and T. Kumbasar, `System Identification and State Estimation with Deep Learning-Supported State-Space Equations'' in \textit{Turkish National Meeting on Automatic Control}, Konya, Türkiye, 2024. 

\bibitem{simba} L. Di Natale, M. Zakwan, B. Svetozarevic, P. Heer, G. Ferrari-Trecate, and C. N. Jones, `Stable linear subspace identification: A machine learning approach,'' in \textit{Proc. Eur. Control Conf.}, Stockholm, Sweden, 2024, pp. 3539--3544. 

\bibitem{simbaext} L. Di Natale, M. Zakwan, P. Heer, G. Ferrari-Trecate, and C. N. Jones, `SIMBa: System identification methods leveraging backpropagation,'' \textit{IEEE Trans. Control Syst. Technol.}, vol. 33, no. 2, pp. 418--433, 2025. 

\bibitem{van1994n4sid} P. Van Overschee and B. De Moor, `N4SID: Subspace algorithms for the identification of combined deterministic-stochastic systems,'' \textit{Automatica}, vol. 30, pp. 75--93, 1994.

\bibitem{MilCal:13} D. N. Miller and R. A. de Callafon, `Subspace identification with eigenvalue constraints,'' \textit{Automatica}, vol. 49, no. 8, pp. 2468--2473, 2013. 

\bibitem{daisy} B. De Moor, P. De Gersem, B. De Schutter, and W. Favoreel, `DAISY: A database for identification of systems,'' \textit{Automatica}, vol. 38, no. 3, pp. 4–5, Sept. 1997. 

% \bibitem{adam} D. P. Kingma and J. Ba, `Adam: A method for stochastic optimization,'' in \textit{Proc. Int. Conf. Learn. Representations}, 2014. 


























\end{thebibliography}
\end{document}